\documentclass[a4paper,11pt]{article}
\usepackage{pos}
\usepackage{gensymb}
\usepackage{caption}
\usepackage{float}

\title{Baikal-GVD Real-Time Data Processing and Follow-Up Analysis of GCN Notices}
\ShortTitle{Baikal-GVD Follow-Up}

\author[a]{V.M.~Aynutdinov}
\author[b]{V.A.~Allakhverdyan}
\author[a]{A.D.~Avrorin}
\author[a]{A.V.~Avrorin}
\author[c,d]{Z.~Barda\v{c}ov\'{a}}
\author[b]{I.A.~Belolaptikov}
\author[a]{E.A.~Bondarev}
\author[b]{I.V.~Borina}
\author[e]{N.M.~Budnev}
\author[l]{V.A.~Chadymov}
\author[f]{A.S.~Chepurnov}
\author*[b,g]{V. Y. Dik}
\author[a]{G.V.~Domogatsky}
\author[a]{A.A.~Doroshenko}
\author[c]{R.~Dvornick\'{y}}
\author[e]{A.N.~Dyachok}
\author[a]{Zh.-A.M.~Dzhilkibaev}
\author[c,d]{E.~Eckerov\'{a}}
\author[b]{T.V.~Elzhov}
\author[d]{L.~Fajt}
\author[l]{V.N. Fomin}
\author[e]{A.R.~Gafarov}
\author[a]{K.V.~Golubkov}
\author[b]{N.S.~Gorshkov}
\author[e]{T.I.~Gress}
\author[h]{K.G.~Kebkal}
\author[a]{I.V.~Kharuk}
\author[b]{E.V.~Khramov}
\author[b]{M.M.~Kolbin}
\author[i]{S.O.~Koligaev}
\author[b]{K.V.~Konischev}
\author[b]{A.V.~Korobchenko}
\author[a]{A.P.~Koshechkin}
\author[f]{V.A.~Kozhin}
\author[b]{M.V.~Kruglov}
\author[j]{V.F.~Kulepov}
\author[e]{Y.E.~Lemeshev}
\author[a,\dagger]{M.B.~Milenin}
\author[e]{R.R.~Mirgazov}
\author[b]{D.V.~Naumov}
\author[f]{A.S.~Nikolaev}
\author[a]{D.P.~Petukhov}
\author[b]{E.N.~Pliskovsky}
\author[k]{M.I.~Rozanov}
\author[e]{E.V.~Ryabov}
\author[a]{G.B.~Safronov}
\author[b,g]{D.~Seitova}
\author[b]{B.A.~Shaybonov}
\author[a]{M.D.~Shelepov}
\author[a]{S.D.~Shilkin}
\author[f]{E.V.~Shirokov}
\author[c,d]{F.~\v{S}imkovic}
\author[b]{A.E.~Sirenko}
\author[f]{A.V.~Skurikhin}
\author[b]{A.G.~Solovjev}
\author[b]{M.N.~Sorokovikov}
\author[d]{I.~\v{S}tekl}
\author[a]{A.P.~Stromakov}
\author[a]{O.V.~Suvorova}
\author[e]{V.A.~Tabolenko}
\author[b]{B.B.~Ulzutuev}
\author[b]{Y.V.~Yablokova}
\author[a]{D.N.~Zaborov}
\author[b]{S.I.~Zavyalov}
\author[b]{D.Y.~Zvezdov}

\affiliation[a]{Institute for Nuclear Research, Russian Academy of Sciences, Moscow, 117312, Russia}
\affiliation[b]{Joint Institute for Nuclear Research, Dubna, 141980, Russia}
\affiliation[c]{Comenius University, 81499 Bratislava, Slovakia}
\affiliation[d]{Czech Technical University in Prague, Institute of Experimental and Applied Physics, 11000 Prague, Czech Republic}
\affiliation[e]{Irkutsk State University, Irkutsk, 664003, Russia}
\affiliation[f]{Skobeltsyn Institute of Nuclear Physics, Moscow State University, Moscow, 119991, Russia}
\affiliation[g]{Institute of Nuclear Physics of the Ministry of Energy of the Republic of Kazakhstan, Almaty, 050032, Kazakhstan}
\affiliation[h]{LATENA, St. Petersburg, 199106, Russia}
\affiliation[i]{INFRAD, Dubna, 141981, Russia}
\affiliation[j]{Nizhny Novgorod State Technical University, Nizhny Novgorod, 603950, Russia}
\affiliation[k]{St.~Petersburg State Marine Technical University, St.~Petersburg, 190008, Russia}
\affiliation[l]{Free researcher, Moscow, Russia}

\emailAdd{viktoriya.ya9@gmail.com}

\abstract{
	The Baikal-GVD alert system was launched at the beginning of 2021. There are alerts for muon neutrinos (long upward-going track-like events) and all-flavour neutrinos (high-energy cascades). The system is able to get a preliminary response to external alerts with a temporal delay of about 3--10 minutes. The Baikal-GVD data processing and the results of the follow-up procedure are described. We report on the analysis of the coincidence in time and direction between the Baikal-GVD cascade GVD20211208CA with an estimated energy of 43 TeV and the announced alert IceCube211208A possibly associated with a flaring state of the blazar PKS 0735+178. }

\ConferenceLogo{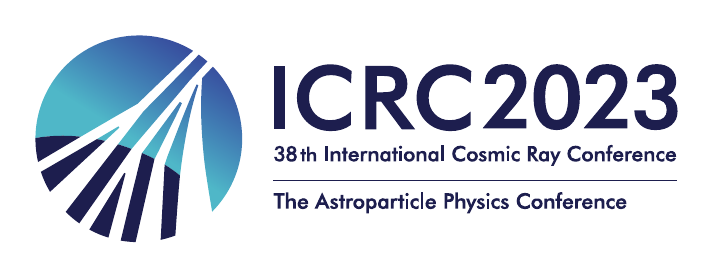}

\FullConference{%
38th International Cosmic Ray Conference (ICRC2023)\\
  26 July -- 3 August 2023\\
  Nagoya, Japan}

\begin{document}
\maketitle

\section{Introduction}

The recent development of multi-messenger astronomy has opened up new avenues for scientific discoveries. It enables simultaneous analyses of data from multiple messengers related to a single astrophysical event.

In 2017, a neutrino event was detected by the IceCube Neutrino Observatory and associated with the blazar TXS 0506+056~\cite{TXS}. This neutrino event coincided in time with gamma-ray emission from the blazar. In 2021, the Baikal-GVD neutrino telescope discovered a high-energy event originating from the direction of TXS 0506+056 which was tentatively associated with a radio flare of that source ~\cite{GVD-TXS}. In addition, the first observation of the diffuse cosmic neutrino flux by Baikal-GVD with a significance of 3.05$\sigma$, is consistent with the results obtained by IceCube \cite{GVD-Diffuse2022}. These studies emphasize the importance of the interplay between neutrino telescopes and other multi-messenger tools.

One significant aspect of multi-messenger astronomy is the ability of experiments to rapidly and efficiently detect astrophysical events, exchange information about them and follow them up with low latency. Collaboration platforms, like the Gamma-Ray Coordinates Network (GCN), enable observatories to share and receive alert messages regarding potential signals from astrophysical sources \cite{GCN}. The latest updates of GCN, which are based on using the new Apache Kafka streaming service, have made alert exchange highly-available to all those interested.

Baikal-GVD is a Cherenkov neutrino telescope situated in Lake Baikal at a depth of 750 to 1275 m. The modular structure of the experiment allows for an increase in volume every year without interrupting the data collection and processing. The current effective volume of the telescope exceeds 0.5 $km^3$. Similar to IceCube and KM3NeT, Baikal-GVD can register neutrinos via detecting products of their interaction - long track-like events and electromagnetic or hadronic cascades. Since 2021, the Baikal-GVD alert system has been upgraded to reduce time delays between detection an event and delivery the information about it. 

\section{Architecture of the Baikal-GVD real-time data processing}

The basic Baikal-GVD structural unit is the optical module (OM), a pressure-resistant glass sphere with a photomultiplier (PMT) inside which converts Cherenkov light emitted by secondary particles during neutrino interactions into waveform signals. Its purpose is to convert the Cherenkov light emitted by secondary neutrino interaction particles into waveform signals. 36 OMs are arranged along a vertical string anchored to the lake floor. These OMs are part of a larger cluster consisting of 8 strings. Currently, Baikal-GVD consists of 13 clusters with each cluster functioning as a standalone detector.

The Baikal-GVD Collaboration has developed a comprehensive data processing system to efficiently process and analyse the experimental data. The system includes several stages, starting from the determination of initial PMT waveforms and ending with the preliminary analysis of the results. The set of C++ programs forming Baikal Analysis and Reconstruction Software (BARS) follows a processing chain where output data from one program serve as the input for the next program.
The BARS programs are managed by the PyBARS workflow which is built on the Luigi Python package \cite{GVD-autom}. 

The raw collected data are transmitted to the storage on the Baikal shore via fibre-optic cable. The data are immediately transferred via a radio channel to the nearest town, Baikalsk. From there it is transmitted through the Internet to the storage facility at the Joint Institute for Nuclear Research in Dubna. At this facility, the data files are processed and reconstructed using a network of virtual machines.

\begin{figure}[htbp]
	\centering 
    \includegraphics[width=1.\textwidth]{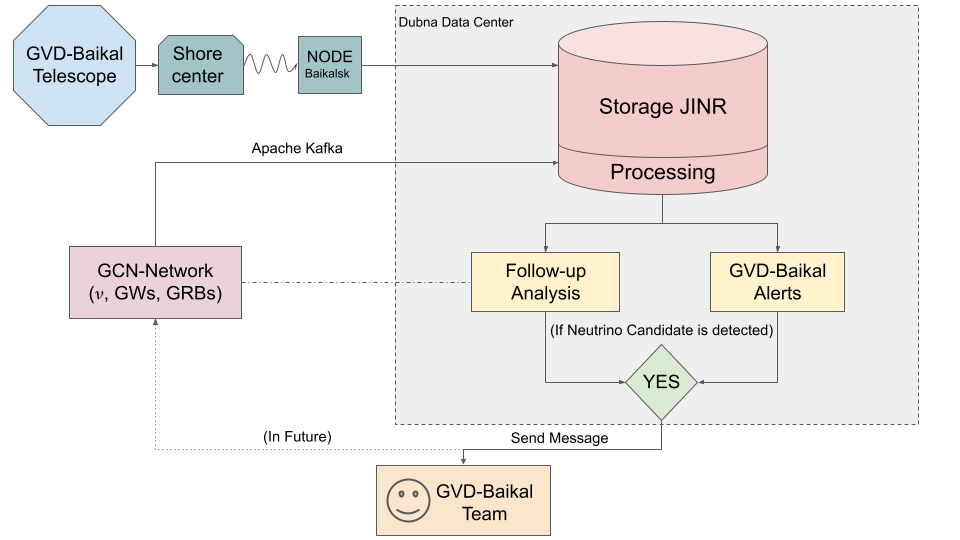}
	\caption{\label{fig:proc} Baikal-GVD data processing chain.}
\end{figure}

After event reconstruction, two alert pipelines are automatically executed (Figure~\ref{fig:proc}). The first pipeline applies a series of selection criteria to filter out background atmospheric muons from the reconstructed cascade-like and track-like events while retaining astrophysical neutrino candidates. The team utilized the Monte Carlo (MC) method to pre-develop all the criteria (see also Section 3). When a program finds a potential neutrino event, an alert notice about the event is sent to the Baikal-GVD team via e-mail. This message includes important parameters, such as the event category, celestial coordinates, timestamp, energy, signalness, background level and service information. The final results are also stored in both the local MySQL database and the file system, in the text format. 

In the second pipeline, machine readable GCN notices are received and stored in both the Influx database and the file system. The data from the storage are used as an input for the follow-up search at the final stage of processing. If there is an external alert event in the database within the selected time windows of Baikal-GVD events, a prompt search for coincidences between celestial coordinates is conducted in the corresponding time windows. The cone and time window of the search depend on the type of external alerts and angular resolution of experiments. Finally, Baikal-GVD team members receive an e-mail containing the search results, including the parameters of both the external (GCN) and the Baikal-GVD event.

Two processing systems are used for event reconstruction and selection. The first system, \textit{Fast Processing}, does not include information about data quality. It reconstructs event coordinates with a lower precision. However, it allows for a reduction in the time delay to 3--5 min between detecting a potential neutrino event and generating a preliminary notice about the candidate. \textit{Offline processing} generates comprehensive information about the event and provides more accurate events coordinates. It requires more time - 24 hours for the run and 2--4 hours for processing.
%, including a run duration of 24 h and an addition 2--4 h for processing.  

\section{Strategy for the real-time neutrino candidate search with Baikal-GVD}

Since 2017, following up astrophysical alerts is one of the tasks of the Baikal-GVD telescope. The gravitational wave event GW170817 resulting from a merger of a binary neutron star and the high-energy astrophysical neutrino IC170922A, originating from the bright blazar TXS 0506+056 were the first events followed up by Baikal-GVD in the offline regime \cite{simc_mm}.

The strategy for searching correlations in the current automatic online processing is based on the following steps which are similar to the offline alert search method. The reconstruction of cascade coordinates involves $\chi^{2}$ minimization of OM hits using their timestamps to obtain the shower vertex coordinates. Then, the maximum-likelihood method is applied to reconstruct the shower energy and the polar and azimuth angles of the direction using the OM amplitude information and reconstructed coordinates \cite{casc_cuts}.
To obtain track parameters, first, the hits associated with a track are identified. Then, time and coordinates of the PMT pulses with the largest charge deposition are obtained. The quality function minimization procedure is used for muon trajectory reconstruction. The quality function depends on time residuals with respect to the direct Cherenkov light from the muon, the sum of charges in OMs, and the distance from an OM to the track \cite{track_epjc}. The muon energy estimation is typically achieved by calculating the average energy loss of muons ~\cite{track_cuts}. 

For such large volume detectors, like Baikal-GVD and IceCube, the angular resolution of the muon direction reconstruction is better than 1$^\circ$ ~\cite{track_cuts},~\cite{ic-tracks}. The accuracy of the cascade direction resolution depends more on optical parameters of the environment. For Baikal-GVD, it is 2--4$^\circ$ ~\cite{casc_cuts}, while for IceCube, it is 10--15$^\circ$ ~\cite{ic-casc}.

A boosted decision tree (BDT) classifier is applied to select neutrino events from the reconstructed events which are primarily background events. If the method selects only upward-going events, there may still be misreconstructed background muons among them. The BDT for cascades was successfully trained with MC samples using a set of variables, including nTrackHits (the number of hits originating from the muon), BranchRatio (the ratio of the number of OM hits above the reconstructed cascade vertex position to that below the vertex), the value of $\chi^2$ after the final cascade position reconstruction, the reconstructed zenith angle, and others \cite{casc_ecrs}. The BDT for tracks incorporates more variables than for cascades, including the $N_{hits}$, $N_{strings}$, maximum distance between OM projections on the track, maximum gap between OM projections along the track, and other variables. It was tested with the low-energy MC atmospheric muon and neutrino samples, demonstrating its high efficiency ~\cite{track_cuts}. Currently, the BDT for high-energy tracks is being developed. 

\begin{figure}[htbp]
	\centering 
	\includegraphics[width=0.9\textwidth]{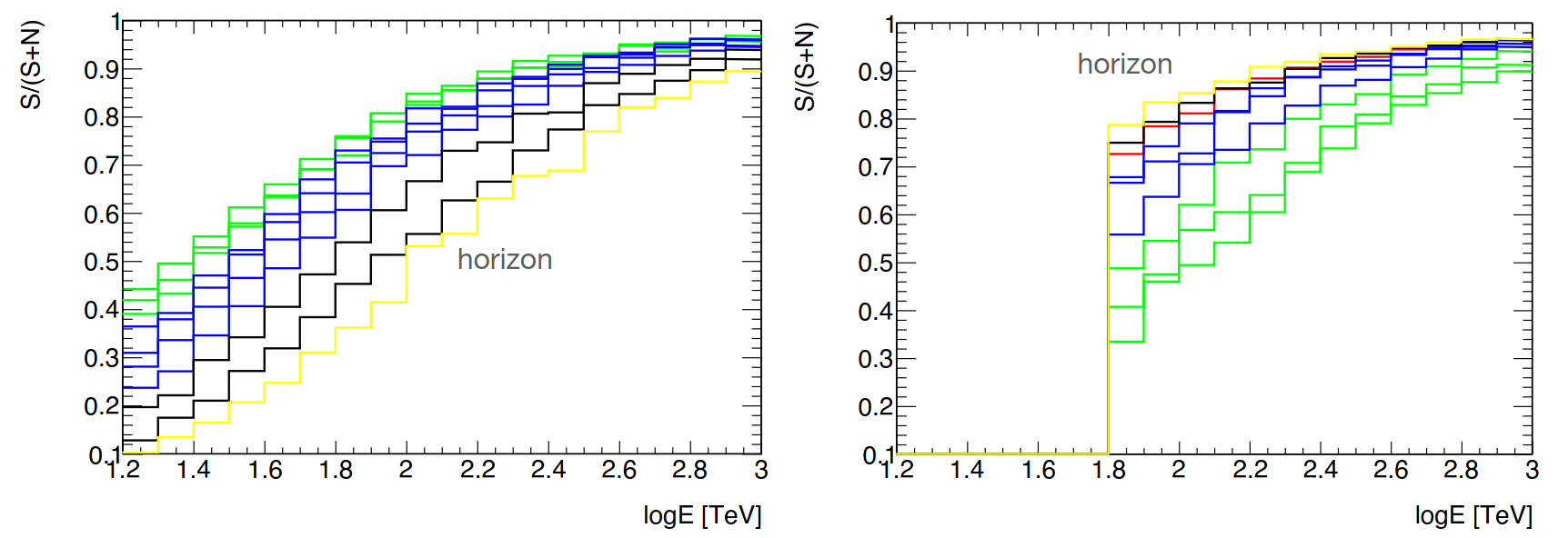}
	\caption{\label{fig:sign} Estimated signalness for Baikal-GVD cascade events, with colours representing a zenith bin with a step of 10$^{\circ}$. The yellow colour corresponds to a zenith angle of 90--100$^{\circ}$. 
		Left: For under-horizon events. Right: for above-horizon events} 
\end{figure}

%As we continue to improve selection techniques, automatic follow-up search provides information about all possible coincidences with all reconstructed events to ensure that we do not miss a real neutrino candidate. 
During automatic processing, the significance of events is evaluated in the preliminary mode by estimating parameters, such as signalness, the false alarm rate and the p-value of being an astrophysical event. Signalness parameter calculation is defined with a two-dimensional probability map in the cos(Zenith) and lgE planes, which is shown in Figure \ref{fig:sign}:

\begin{equation} \label{eq:sign}
	P_{i,j} = \frac{N^{sig}_{i,j}}{(N^{sig}_{i,j} + N^{bkg}_{i,j})} 
\end{equation}

$N_{sig}$ and $N_{bkg}$ are calculated using MonteCarlo simulations: $N_{sig}$ represents the number of simulated astrophysical events ($E^{-2.46}$) while $N_{bkg}$ is the number of events of atmospheric prompt and conventional neutrinos and atmospheric muons for the given zenith and energy bins. The false alarm rate represents the number of $N_{bkg}$ per year with the same or higher signalness value. 

Additionally, automatic processing can determine the preliminary p-value, which represents the Poisson probability of atmospheric origin of an event for the background-only hypothesis. The background is determined using a time-scrambled data sample of real events from the previous years.  

Offline searches include calculating upper limits on the neutrino fluence $E^2 \cdot \Phi_{\nu}(E)$  for the energies ranging from 1 TeV to 10 PeV assuming a spectrum of $E^{-2}$ or $E^{-2.46}$ and the equal fluence in all flavours. Based on the 2020--2021 Baikal-GVD data, the limits for one cluster were found to be in the range of 1 to 3  $GeV/cm^{2}$~\cite{alerts_ecrs} within a time window of $\pm$ 12 h.

\section{The latest results of the Baikal-GVD follow-up analysis}

\begin{figure}[htbp]
	\centering 
	\includegraphics[width=.48\textwidth]{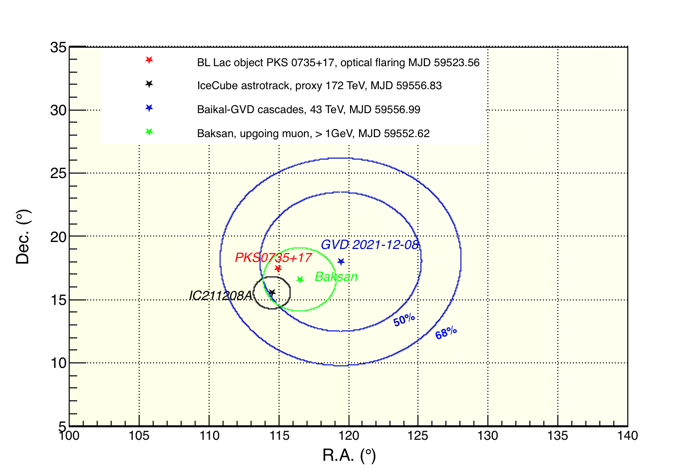}  
	\includegraphics[width=.508\textwidth] {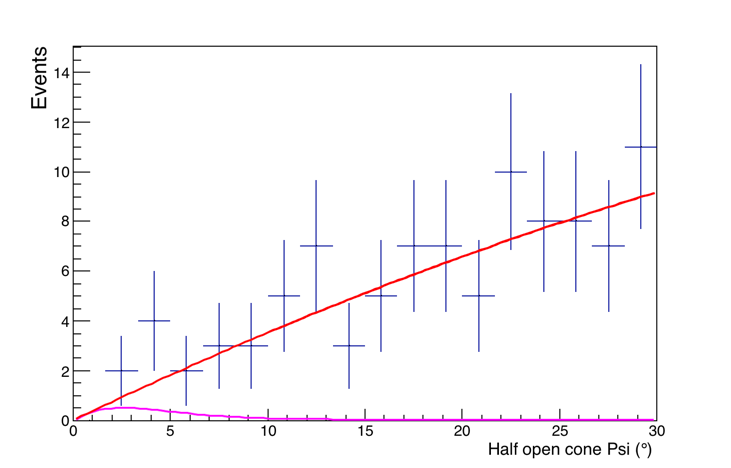}
	\caption{\label{fig:pks} Left: arrival directions of the Baikal-GVD, IceCube and Baksan candidates, along with their directional uncertainties with respect to PKS 0735+17. Right: angular distance distribution to PKS 0735+17 for the observed Baikal-GVD events (blue points with error bars), the expected background (red), and the expected signal events (purple) based of the best-fit spectrum ~\cite{alerts_ecrs}}
\end{figure}

The initial intriguing finding from the follow-up investigation with Baikal-GVD was the discovery of a correlation, both in direction and time, between the Baikal-GVD cascade event GVD211208A (E $\approx$ 43 TeV)~\cite{baikal-Atel} and the up-going track IC211208A (E $\approx$ 172 TeV) with a time difference of 4 h. Both events are located in the direction of a flaring source, PKS 0735+17, which emits in the $\gamma$, X-ray, ultraviolet, and optical bands. Similar to the case of TXS 0506+056, this source is considered a "masquerading BL Lac"~\cite{pks0735}. Additionally, the KM3NeT (E $\approx$ 18 TeV)~\cite{km3net-Atel} and Baksan (E > 1 GeV)~\cite{baksan-Atel} neutrino observatories have announced the observation of neutrino candidates during this flare. The sky region where the source PKS 0735+17 and the events IC211208A, GVD211208A and the Baksan are located is depicted in Figure \ref{fig:pks} (left). 

%Baikal-GVD observed the downward neutrino candidate in 3.95 h after the IC211208A.
The event originated from the direction, which is 4.68$^{\circ}$ away from PKS 0735+17 and 5.30° away from the the best-fit direction of IC211208A. 
%According to Monte Carlo simulations the Baikal-GVD PSF (Point Spread Function) has a 50\% containment radius of 5.5$^{\circ}$ and a 68\% containment radius of 8.1$^{\circ}$.
Within a circle with a radius of 5.5$^{\circ}$ around this source, one can reasonably expect only approximately 0.0044 events to occur per 24 h. The Poisson probability of chance coincidence is 0.0044, which corresponds to a significance of 2.85$\sigma$. To further investigate the flaring source PKS 0735+17, we conducted a follow-up study using the data set collected by Baikal-GVD from April 2018 to November 2021 and the unbinned likelihood method (see Figure \ref{fig:pks}, right) ~\cite{alerts_ecrs}. %For this, the mismatch angle distribution (Figure \ref{fig:pks}, right) is fitted using the following likelihood function:
%\begin{equation}
	%L=\prod_{i=1}^{N_{obs}} \frac{e^{-\lambda_i}\lambda_i^{n_i}}{n_i!};
	%L=\prod^{N_{obs}} Poisson(n^i_{obs}; (\alpha S_{i} + (1-\alpha) B_{i})),
%\end{equation}
%where the free parameter is the relative signal fraction $\alpha$, constrained to values between 0 and 1, while $B_{i}$ and $S_{i}$ are the expected background and signal distribution functions, respectively. Here, we used ON/OFF areas around the source as 10$^\circ$/30$^\circ$ and performed scrambling of RA to estimate background events $B_{i}$. Observed events, background and expected contribution of signal events $S_{i}$ from the best-fit procedure around the PKS 0735+17 direction are shown in Figure \ref{fig:pks} (right). 
The post-trial probability of the background origin is found to be 1.13$\sigma$, which is not considered very significant for the association. The recent research suggests that certain models predict the production of low-energy neutrinos through interactions between protons and dense matter~\cite{pks0735}. These findings could potentially provide an explanation for the results observed in the Baikal-GVD and Baksan experiments. 
%According to FERMI observations, PKS 0735+17 remained active throughout the second half of 2022. Preliminary Baikal-GVD data of 2022 season shows there are no cascade candidates detected in this specific direction above the background level. 

In the online processing for the Season 2022 (March 2022 -- March 2023), no candidates have been found in the Baikal-GVD follow-up searches for temporal and spatial coincidences with IceCube neutrino alerts and bright gamma-ray bursts (GRBs). The preliminary offline data also suggest that there are no coincidences within real-time windows of $\pm$500 s and $\pm$1000 s. According to the offline data, five cascade candidates have been identified in the IceCube tracks follow-up searches in the time window of $\pm$ 24 h. These candidates have a reconstructed energy of more than 60 TeV. The most intriguing candidate was observed for IC230112A occurring 2.5 h prior to the IceCube event. This coincidence was found within a cone of 1.37$^{\circ}$ between the two events. The event was reconstructed as an upward-going event with an energy of E $\approx$ 97.7 TeV. However, the number of event hits did not reach the threshold value used to select events of good quality. 

When IceCube observes the up-going tracks, they often appear as down-going tracks for Baikal-GVD due to the location of the latter. Specifically, within a time window of $\pm$24 h, horizontal coordinates of these alerts do not have a zenith angle greater than 120$^{\circ}$, which almost excludes the presence of neutrino candidates among well-reconstructed low-energy Baikal-GVD up-going track events.
However, one track neutrino candidate was detected in the vicinity of IC220918A which is close to the blazar TXS 0506+056. Notably, this occurrence was registered 15 mins prior to the IceCube event. It is reconstructed as an up-going event (zenith $\approx$ 121$^{\circ}$) with the most probable energy E $\approx$ 141 TeV. The angle between two events is approximately 4$^{\circ}$ while the angle between TXS 0506+056 and the Baikal-GVD muon track is 3.1$^{\circ}$.
However, the following analysis revealed that this event is most likely due to the atmospheric background from muon bundles.
% Meanwhile, the angle with GVD210418CA is 8.02996$^{\circ}$.
% Once an improved version of track reconstruction is implemented, which includes both multicluster and single cluster event reconstruction, we will conduct the search again.

Considering the search for correlation of Baikal-GVD and IceCube cascades, the method for the point follow-up analysis is challenging due to the low angular resolution in this mode. During follow-up searches in 2022, we observed an up-going cascade neutrino candidate, GVD220625CA, in the direction of the IceCube cascade IC220625CA. Both events are high-energy events: the IceCube reconstructed energy is 481.28 TeV while Baikal-GVD registered an event with E $\approx$ 144.5 TeV. The signalness parameter was calculated according to Figure \ref{fig:sign}. The probability of the astrophysical origin for this event is higher than 80$\%$. For the IceCube, the false alarm rate is 0.3 while for Baikal-GVD it is 0.08 atmospheric background events per year. The preliminary p-value for GVD220625CA as to being the atmospheric background event is 0.1\%, which corresponds to a significance level of 3.3$\sigma$. This coincidence is interesting for a future comprehensive research, but it goes beyond the scope of the paper.

Baikal-GVD also has been following up bright gamma-ray bursts which may be caused by the collapse of massive stars or the merging of compact binary systems. Due to their significant energy output, they are considered as potential sources of ultra high-energy cosmic rays and neutrinos. The Baikal-GVD processing system conducts a prompt search for neutrino sources 24 h before and after gamma-ray bursts (GRBs). Like many other experiments, Baikal-GVD also followed up the brightest burst of GRB 221009A. One up-going track neutrino candidate has been registered by Baikal-GVD at the angle $\psi$ = 2.76$^{\circ}$ with respect to GRB 221009A, 9 h after the Fermi Gamma-Ray Space Telescope announced the event. The most probable value of the reconstructed energy is approximately 36 TeV. The event was rejected due to a high probability of being a misreconstructed background event. 

\section{Conclusion}

The automatic processing of the Baikal-GVD data has been efficiently functioning since 2021, with the ability to work in a quasi-online regime. Baikal-GVD receives GCN notices via the Apache Kafka protocol. This allows for the real-time study of multi-messenger events in real time and is a valuable addition to the IceCube and KM3NeT observatories in neutrino sky observation. The follow-up analysis of IceCube tracks and the bright GRB 221009A in $\pm$500 s, $\pm$ 1h, $\pm$ 24 h did not reveal Baikal-GVD neutrino candidates in the directions of the alerts in 2022. In addition to the previously reported coincidence with PKS 0735+17 in 2021, we presented a high-energy neutrino candidate in the direction of the cascade IC220625CA in 2022. 
%Mainly, Baikal-GVD focuses on cascade-like events, which is beneficial for conducting a more comprehensive analysis, such as the discovery of diffuse flux. In addition to well-known coincidence with PKS0735+17 in 2021 year, we observed a high energy neutrino candidate in direction of IC220625CA in 2022. 

%The work of V. Dik was supported by grant of JINR for young scientists № 23-202-02. 
This work is partially supported by the European Regional Development Fund--Project "Engineering applications of microworld physics" (CZ02.1.01/0.0/0.0/16 019/0000766).

%% Full authors list (ONLY FOR COLLABORATIONS)
%\clearpage
%\section*{Full Authors List: \Coll\ Collaboration}
%
%\noindent \textbf{Note comment afterwards:} Collaborations have the possibility to provide an authors list in xml format which will be used while generating the DOI entries making the full authors list searchable in databases like Inspire HEP. \\
%
%\scriptsize
%\noindent
%first.author$^1$, 
%second.author$^2$, 
%third.author$^3$ % .... more names
%and 
%last.author$^{n}$ \\
%
%\noindent
%$^1$first.affiliation.
%$^2$second.affiliation. % .... more affiliation
%$^{m}$last.affiliation.

\end{document}